\documentclass[10pt,aps,prb,twocolumn,showpacs,superscriptaddress,floatfix,nofootinbib]{revtex4-2}

\pdfoutput=1
\usepackage[utf8]{inputenc}
\usepackage[english]{babel}
\usepackage[T1]{fontenc}
\usepackage{physics} 
\usepackage{csquotes}
\usepackage{graphicx}

\usepackage{amsmath,amsfonts}

\usepackage[dvipsnames]{xcolor}
\usepackage[colorlinks=true]{hyperref}
\usepackage[capitalise]{cleveref}

\usepackage{listings}

\usepackage{bm} 
\usepackage{xcolor}
\usepackage[normalem]{ulem}

\renewcommand{\vec}[1]{\bm{#1}}

\begin{document}

\title{Accurate ground states of $SU(2)$ lattice gauge theory in 2+1D and 3+1D}

\author{Thomas Spriggs}
\affiliation{QuTech and Kavli Institute of Nanoscience, Delft University of Technology, 2628 CJ, Delft, The Netherlands}
\author{Eliska Greplova}
\affiliation{QuTech and Kavli Institute of Nanoscience, Delft University of Technology, 2628 CJ, Delft, The Netherlands}
\author{Juan Carrasquilla}
\affiliation{Institute for Theoretical Physics, ETH Zürich, 8093, Switzerland}
\author{Jannes Nys}
\affiliation{Institute for Theoretical Physics, ETH Zürich, 8093, Switzerland}

\begin{abstract}
We present a neural network wavefunction framework for solving non-Abelian lattice gauge theories in a continuous group representation.
Using a combination of $SU(2)$ equivariant neural networks alongside an $SU(2)$ invariant, physics-inspired ansatz, we learn a parameterization of the ground state wavefunction of $SU(2)$ lattice gauge theory in 2+1 and 3+1 dimensions. Our method, performed in the Hamiltonian formulation, has a straightforward generalization to $SU(N)$. We benchmark our approach against a solely invariant ansatz by computing the ground state energy, demonstrating the need for bespoke gauge equivariant transformations. We evaluate the Creutz ratio and average Wilson loop, and obtain results in strong agreement with perturbative expansions. Our method opens up an avenue for studying lattice gauge theories beyond one dimension, with efficient scaling to larger systems, and in a way that avoids both the sign problem and any discretization of the gauge group.
\end{abstract}

\maketitle
\section{Introduction}

The standard model is currently our best candidate for the fundamental theory of the physical forces of our Universe. Constituent elements of the standard model include matter and gauge fields that are described by quantum field theory. One of these fundamental forces, the strong nuclear force, has been the subject of immense study. To arrive at a regularized, well defined theory, Wilson reformulated the theory of strong interactions from a continuous space-time to one on a discrete, Euclidean lattice where time is imaginary and periodic~\cite{wilson1974confinement}. Soon after, Kogut and Susskind introduced the Hamiltonian formulation which allowed time to remain real and continuous but proved numerically more challenging to solve~\cite{kogut1983lattice}. These lattice gauge theories (LGTs) offer rich avenues for simulating and studying quantum field theories in non-perturbative regimes.

A plethora of numerical techniques emerged to study the \emph{Euclidean} formulation of LGTs; quantum Monte Carlo (QMC)~\cite{creutz1983monte}, one of the most successful approaches to simulating LGTs, has led to a range of new insights into the properties of non-perturbative systems from neutron starts~\cite{baym2018hadrons} to high-energy physics and nuclear physics~\cite{davoudi2022report}, for example by giving insight into the hadronic spectrum using lattice QCD~\cite{briceno2018scattering, edwards2011excited, gross202350}. See Ref.~\cite{gattringer2009quantum} for a pedagogical introduction to LGTs. Within this field, many advancements have been made to improve the efficiency of generating configurations of gauge fields - often the most computationally intensive part of any LGT measurement. These range from the introduction of hybrid Monte Carlo algorithms~\cite{PhysRevD.28.1506,PhysRevLett.51.2257,PhysRevLett.55.2774}, to machine learning approaches like normalizing flows~\cite{kanwar2020equivariant, boyda2021sampling,Abbott:2024kfc, komijani2025normalizing}.
Despite their remarkable success, QMC-based methods face limitations in certain parameter regimes of LGTs, particularly when dealing with finite baryon chemical potentials, topological $\theta$-terms, or out-of-equilibrium phenomena. In such scenarios, the well-known sign problem renders the QMC numerical approach ineffective and unreliable~\cite{nagata2022finite, bauer2023quantum, alexandru2022complex, usqcd2019hot, troyer2005computational,Danzer:2009dk}.

In part guided by the limitation imposed by the sign problem, several methods have thus been applied to the \emph{Hamiltonian} formulation of LGTs.

Tensor networks (TNs) and the density matrix renormalization group (DMRG) algorithm, given their immense progress in the field of condensed matter, have become useful tools to simulate LGTs and extract information about their entanglement structure~\cite{haegeman2015gauging, buyens2014matrix, banuls2018tensor, magnifico2024tensor, silvi2014lattice}. 
These methods have demonstrated significant success in simulating LGTs in 1+1 dimensions for both Abelian and non-Abelian gauge groups~\cite{byrnes2002density, silvi2014lattice, pichler2016real, silvi2019tensor, funcke2020topological, buyens2014matrix, rico2014tensor, haegeman2015gauging, silvi2017finite, buyens2017finite, banuls2017density, ercolessi2018phase, magnifico2019symmetry, sala2018gaussian, banuls2017efficient, rigobello2021entanglement, angelides2023computing, schmoll2023hamiltonian, osborne2023probing, kebrivc2023confinement, belyansky2024high, papaefstathiou2025real, calajo2024quantum}. More recently, they have been applied to Abelian LGTs in up to 3+1 dimensions~\cite{felser2020two, emonts2023finding, magnifico2021lattice, knaute2024entanglement}, and non-Abelian theories with truncated gauge fields in Refs.~\cite{Cataldi_2024} and \cite{balaji2025quantumcircuitssu3lattice} for 2+1D $SU(2)$ and 3+1D $SU(3)$ respectively.

A second class of approaches to the simulation of Hamiltonian LGTs are digital quantum devices~\cite{byrnes2006simulating, martinez2016real, mathis2020toward, paulson2021simulating, kokail2019self, klco20202, a2022self, d2025adiabatic} and analogue quantum simulators~\cite{aidelsburger2022cold, mil2020scalable, yang2020observation, zhou2022thermalization, schweizer2019floquet, gorg2019realization}. These have been used as novel platforms to gain insight into emergent properties of gauge theories~\cite{banuls2020simulating, bauer2023quantum, klco2022standard, mil2020scalable, halimeh2025cold,balaji2025quantumcircuitssu3lattice}.
Recent breakthroughs include the realization of discrete gauge symmetries and the observation of gauge-invariant phenomena like confinement and string-breaking dynamics in Abelian and non-Abelian theories~\cite{aidelsburger2022cold, schweizer2019floquet, gorg2019realization, zohar2022quantum}. Particularly significant is the demonstration of real-time dynamics in $\mathbb{Z}_2$ and $U(1)$ lattice gauge theories, allowing controlled studies of thermalization, confinement dynamics, and the formation of topological defects~\cite{yang2020observation, zhou2022thermalization}.
Quantum simulations of non-Abelian lattice gauge theories, such as $SU(2)$ and $SU(3)$ in 1+1D, have been explored on small lattices~\cite{farrell2023preparations, atas20212, farrell2023preparations2}. Limited simulations exist for 2+1D~\cite{ciavarella2022preparation} and 3+1D is only recently being explored~\cite{balaji2025quantumcircuitssu3lattice}.

One of the challenges facing these first two methods is the infinite nature of either the continuous gauge group or electric field spectrum; these methods often require discretizing the group into a subgroup or truncating the gauge fields. These digitization methods can struggle at increasing spatial dimensions and certain regimes of LGTs~\cite{k9p6-c649, magnifico2024tensor}.

One further approach is variational Monte Carlo (VMC). Early applications of VMC to LGTs relied on hand-crafted wavefunction parametrizations, often stemming from perturbative expansions of the theory~\cite{chin1985exact, chin1988exact,HEYS198519,Arisue:1982tt,HUANG1988733}. The limited wavefunction parametrization has historically posed a limitation to VMC accuracies. Modern applications of VMC have overcome these limitations through the use of physics-agnostic, neural network parameterizations of the wavefunction which has achieved impressive, and often state-of-the-art, results in the field of condensed matter~\cite{carleo2017solving, chen2024empowering, viteritti2023transformer, Rende_2025, pescia2024message, kim2024neural, linteau2024phase, robledo2022fermionic, medvidovic2024neural, Spriggs_2025, Hernandes:2025buh, wu2024variational}, quantum chemistry~\cite{pfau2024accurate, pfau2020ab, qian2024deep, gao2023generalizing, hermann2020deep, scherbela2023towards, li2022ab}, and nuclear and high-energy physics~\cite{fore2023dilute, gnech2024distilling, rinaldi2022matrix}. The application of these \emph{neural wavefunctions} to LGT is limited to Abelian gauge theories: $\mathbb{Z}_2$ ~\cite{luo2021gauge,PhysRevB.110.165133} and $U(1)$~\cite{luo2022gauge}.

We present a scalable neural network wavefunction framework for simulating non-Abelian
lattice gauge theories free of any discretizations of the gauge fields as well as the sign problem. Whilst the approach is general for any $SU(N)$ theory, we focus on $SU(2)$ in 2+1 and 3+1 dimensions. In order to respect gauge invariance whilst allowing for an expressive ansatz to represent the wavefunction, we introduce the lattice gauge neural wavefunction (LGNWF) ansatz, parameterized by expressive neural networks, containing equivariant and invariant blocks, well suited for VMC. To demonstrate the power of this ansatz we show that finds ground states with lower energy than a prototypical physics-inspired ansatz and then present observables in 2+1 and 3+1 dimensions that agree with QMC calculations on similar sized lattices. Our framework allows for simple extensions to larger systems, higher dimensions, and different gauge groups; there is also a path to explore the inclusion of fermions, real-time evolution, and non-equilibrium dynamics.

\section{Formalism}

\subsection{Hamiltonian and representation} \label{sec:YM_Hamiltonian}
We focus on $SU(2)$ pure gauge theory, where the gauge fields are defined on the links between spatial lattice sites. For a given lattice site $x$ the directed link connecting $x$ with its neighbor in the direction $\mu$ is denoted $U_\mu(x)$. These are related to the gauge fields from the continuum theory, $A_\mu^a(x)$, through $U_\mu(x) = \exp(- i \frac{1}{2} \sum_a\sigma^a A_\mu^a(x))$, with $a=1,2,3$ and $\sigma^a$ being the Pauli matrices. An entire lattice of links will be denoted $\vec{U}$.

The Kogut-Susskind formulation of the $SU(2)$ LGT Hamiltonian is given by~\cite{chin1985exact, kogutsusskind1975hamiltonian}
\begin{equation}
    H = \frac{g^2}{a_s} \left( \frac{1}{2}\sum_{l,a} E_l^a E_l^a + \lambda \sum_{x, \nu>\mu} \left[1 - \frac{1}{2}\text{Tr} P_{\mu,\nu}(x)\right] \right).
    \label{eq:hamiltonian_abstract_form}
\end{equation}
Here, $l$ indexes each link in the lattice, and the sum over $\mu,\nu,x$ encompasses each plaquette on the lattice. Moreover, $g$ is the coupling constant, $\lambda = 4/g^4$, and $a_s$ is the lattice spacing. Throughout this work, we will use a rescaled Hamiltonian instead, $\frac{a_s}{g^2}H$, see Appendix~\ref{app:sec:Hamiltonian} for more details. The (untraced) plaquette $P_{\mu,\nu}(x)$ is the product of four link operators forming a closed counter-clockwise loop whose bottom left node is $x$. The electric field operators $E^a$ are either the left, or right, generators of $SU(2)$. They are defined through the commutators $[E^a,U] = \frac{1}{2}\sigma^aU$ or $[E^a,U] = U\frac{1}{2}\sigma^a$ for the left or right generators respectively. The gauge links transform as $\mathcal{T}_\Omega U_{\mu}(x) = \Omega(x) U_{\mu}(x) \Omega(x+\mu)^\dagger$, where $\Omega \in SU(2)$. The theory is symmetric under $SU(2)$ gauge transformations, meaning our wavefunction is required to be invariant $\Psi(\mathcal{T}_\Omega\vec{U}) = \Psi(\vec{U})$.

There are two features to this theory that prove difficult for contemporary quantum simulation methods (such as quantum simulations and TNs): representing the continuous nature of the gauge group and respecting the gauge symmetry. 

The continuous nature has several workarounds that involve some level of digitization of the gauge fields. One approach is the employ the quantum link model formulation~\cite{HORN1981149,ORLAND1990647,PhysRevD.60.094502,CHANDRASEKHARAN1997455,PhysRevResearch.2.023015,byrnes2006simulating,PhysRevResearch.3.043209,bauer2023quantum} where the representation is discrete but the link variables are no longer unitary - a relic that must be removed before this model can be compared with the original theory. Alternatively, remaining with the Kogut-Susskind form, one can construct explicit basis representations known as the electric or magnetic bases (see Ref.~\cite{PhysRevD.104.074505} for an overview of bases in 1+1D $SU(2)$). The former basis involves constructing a basis of a finite number of irreducible representations (irreps) of the electric field operator, relying on an explicit truncation on the number of irreps to keep the basis finite; however, this truncation is known to be increasingly less reliable at smaller values of $g$, when the dimensionality of the system increases, or when trying to approach the continuum limit of the theory~\cite{k9p6-c649, magnifico2024tensor}. The latter, the magnetic basis, involves direct parameterizations of the group elements of the theory: this is the approach taken in this work. For implementations on quantum simulators or TNs the magnetic basis is truncated either by sampling the continuous group or using a discrete subgroup~\cite{ercolessi2018phase,magnifico2020real}. In our work no such restrictions are added and we work with the full, continuous gauge group.

As for respecting the gauge symmetry, several works are formulated in a basis that is gauge invariant by construction. These go by the name of dressed site formulations in TNs for which we refer the reader to Ref.~\cite{magnifico2024tensor}.
In the context of VMC, Ref.~\cite{vieijra2020restricted} showed that the states of $SU(N)$ can be represented using a discrete spin-$j$ angular momentum representation, which has dimension $2j+1$. Using the straightforward representation of the symmetry in this basis, the model can be made global $SU(N)$ symmetric by construction. This was then extended to \emph{local} $SU(N)$ symmetry~\cite{luo2023gauge}. Nevertheless, as pointed out in Ref.~\cite{vieijra2021many}, working only with these gauge-invariant bases severely limits the tractability in higher dimensions and the expressive power of the neural networks.

Our approach to representation works with the continuous gauge group without any discretization on truncation at any point. We achieve this by writing Eq.~\eqref{eq:hamiltonian_abstract_form} not in terms of operators acting on \textit{elements} of $SU(2)$ but in terms of the \textit{algebra} of the group instead, as was originally done in Ref.~\cite{chin1985exact}. This leads to a continuous theory in $S^3$ that can be tackled using VMC. Furthermore, as we will see in Section~\ref{sec:nnmodel} we can create a gauge invariant trial wavefunction that exactly respects the symmetry without resorting to a restricted basis. 

To be explicit, we use a spherical representation for the link variables $(\rho, \theta, \phi)$,
\begin{equation}
    U_\mu(x) = \cos\left(\frac{\rho}{2}\right) \mathbb{I} - i n \cdot\sigma\sin\left(\frac{\rho}{2}\right),
    \label{eq:link_ito_polars}
\end{equation}
where $n = (\sin\theta\cos\phi, \sin\theta\sin\phi, \cos\theta)$ given $0\leq \rho \leq 2\pi$, $0\leq \theta \leq \pi$, and $0\leq \phi \leq 2\pi$. In this representation, only the electric field term of the Hamiltonian changes, it becomes the Laplace-Beltrami operator on $S^3$
\begin{equation}\label{eq:EE}
    E^2 = -\frac{1}{4}\nabla^2_{S^3}.
\end{equation}
Further details relating to the spherical representation are given in Appendix~\ref{sec:impl_details} and \ref{app:sec:Lapl_belt}.

\subsection{Variational model}\label{sec:nnmodel}

As previously stated, our variational wavefunction needs to be gauge invariant, and whilst neural networks can approximate a rich class of functions, and thus one could possibly learn the gauge symmetry through training, we take the approach of explicitly baking-in gauge symmetry into our variational model. We will now present two ways that this can be achieved.

The simplest approach is to work with only gauge invariant constructions. The smallest of which is the traced plaquette
\begin{align}
    W_{\mu, \nu}(x) &= \frac{1}{2}\text{Tr}{P_{\mu, \nu}(x)}.
\end{align}
The first ansatz we consider does not contain a neural network, instead, traced plaquettes are combined to form a physics-informed, two-body Jastrow ansatz~\cite{PhysRev.98.1479}, the output of which remains invariant under any gauge transformation of the input
\begin{align}
    \Psi(\vec{U}) &= \prod_{\mu < \nu, x} e^{\alpha_\mu W_{\mu, \nu}(x)} \prod_{\mu' < \nu', x'}  e^{\beta_{ d(x, x')}^{\mu, \nu, \mu', \nu'} W_{\mu, \nu}(x) W_{\mu', \nu'}(x')}.\label{eq:plaquette_model}
\end{align}
Here, $\alpha$ and $\beta$ are complex variational parameters, where the latter depends on the relative distance $d(x, x')$ between sites $x$ and $x'$, yielding a translationally invariant, yet non-local model. Setting $\beta = 0$ yields the single-plaquette model from Refs.~\cite{chin1985exact, chin1988exact,Arisue:1982tt,HUANG1988733}.

Extending beyond this, recent work focusing on using neural networks for sampling gauge configurations has demonstrated that it is feasible to construct gauge equivariant neural networks for lattice gauge theories using continuous group representations~\cite {favoni2022lattice, kanwar2020equivariant, boyda2021sampling}. This offers a powerful approach to capturing correlations in lattice gauge theories, and offers a promising extension to tackle continuous-group theories beyond one dimension. We now introduce the LGNWF ansatz, comprised of two such gauge equivariant neural networks.

In our first type of neural transformation, we update a link $U_\mu(x)$ through updating the eigenvalues of an untraced plaquette operator $P_{\mu, \nu}(x)$ (which contains $U_\mu(x)$). This is done using a neural network that takes all traced plaquette loops $W$ as an input, i.e.\ 
\begin{align}
    P_{\mu, \nu}(x) &= V e^{i \lambda} V^\dagger \to P'_{\mu, \nu}(x) = V f(\lambda,\vec{W}) V^\dagger  ,  
    \label{eq:link_equivariant_equations-1}
    \\
    U_\mu(x) &\to U'_\mu(x) = P'_{\mu,\nu}(x) P_{\mu,\nu}(x)^\dagger U_\mu(x),
    \label{eq:link_equivariant_equations-2}
\end{align}
where $\lambda = [\lambda_1, \lambda_2]$ is a vector of the sorted eigenvalues of $P$, $V$ are the corresponding eigenvectors, and $f$ is a convolutional neural network (CNN) architecture taking all invariant traced plaquettes as input. The transformation in Eqs.~\eqref{eq:link_equivariant_equations-1}--\eqref{eq:link_equivariant_equations-2} was introduced in~\cite{kanwar2020equivariant} and is referred to here as a \textit{plaquette equivariant layer}. The second neural transformation combines all untraced plaquettes that start and end at $x$, i.e.\ $C_{\mu, \nu}^i(x) = \{P_{\mu,\nu}(x), P_{\mu,-\nu}(x), P_{-\mu,\nu}(x),  P_{-\mu,-\nu}(x)\}^i$. We then parametrize a transformation of $U_\mu(x)$ as 
\begin{align}
    \epsilon_\mu(x) &= e^{i\sum_{i} \gamma_{\mu,i} [C_{\mu, \nu}^i(x)]_{aH}}  
    \label{eq:node_equivariant_equations-1}
    \\
    U_\mu(x) &\to U'_\mu(x) = \epsilon_\mu(x) U_\mu(x),
    \label{eq:node_equivariant_equations-2}
\end{align}
where $\gamma_\mu$ are variational parameters for the transformation along direction $\mu$, and $aH$ is shorthand for the anti-Hermitian component; this transformation was introduced in Ref.~\cite{favoni2022lattice}. We refer to this transformation as a \textit{node equivariant layer}. 

\begin{figure}
    \centering
    \includegraphics[width=\linewidth]{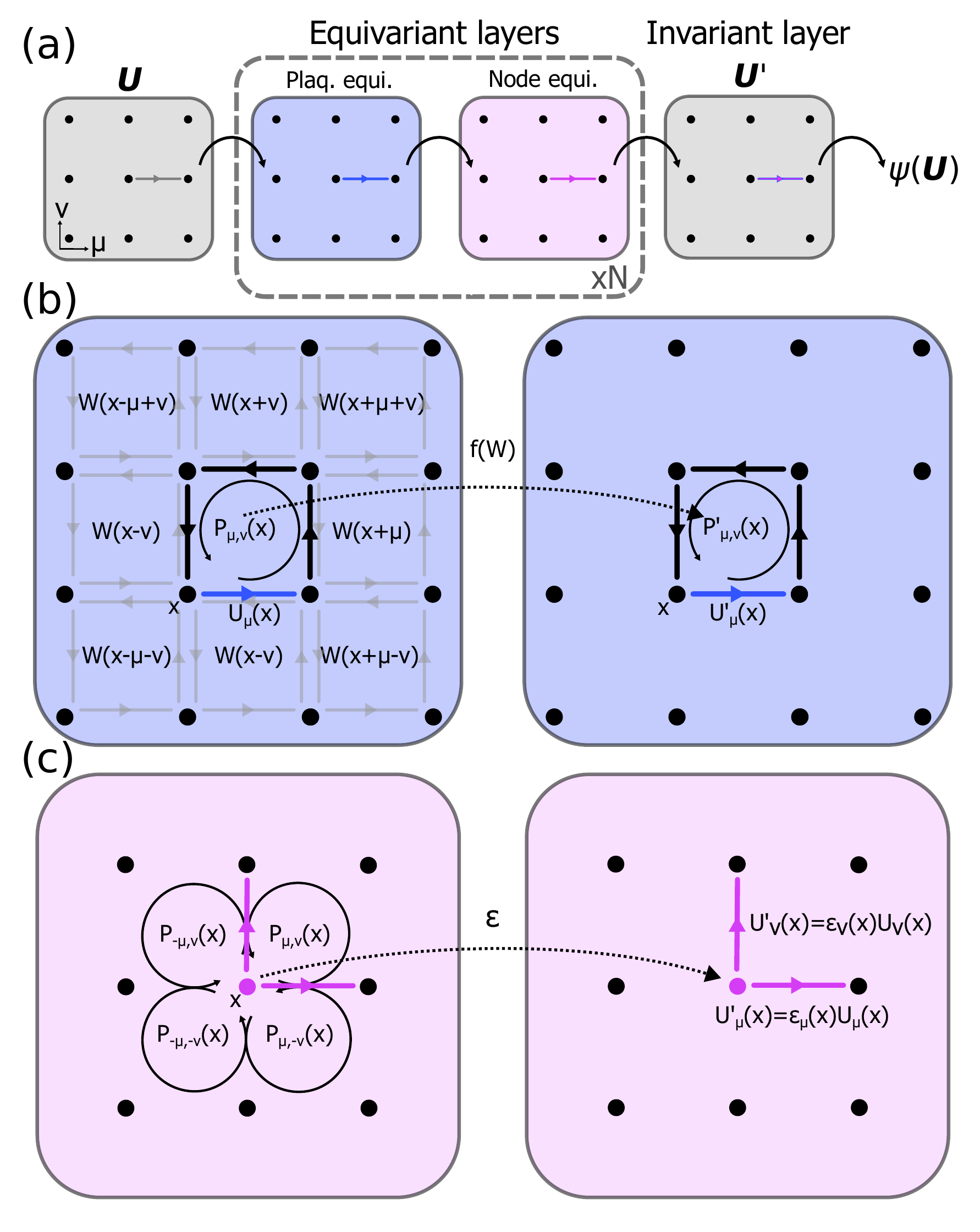}
    \caption{Example configuration of the LGNWF ansatz. (a) Schematic of updating a lattice of links, $\vec{U}$, through $N$ blocks of equivariant layer-to-layer transforms (b) and (c). (b) \textit{Plaquette equivariant network} updating the link $U_{\mu}(x)$ through Eqs.~\eqref{eq:link_equivariant_equations-1} and \eqref{eq:link_equivariant_equations-2} conditioned on non-local information $W$. (c) \textit{Node equivariant layer} acting on $U_{\mu}(x)$ and $U_{\nu}(x)$ through Eqs.~\eqref{eq:node_equivariant_equations-1} and \eqref{eq:node_equivariant_equations-2} acting only on plaquettes originating at $x$.}
    \label{fig:architecture}
\end{figure}

We capture higher-order correlations through these gauge-equivariant neural transformations on the configurations $\vec{U}$, before entering them into the invariant form in Eq.~\eqref{eq:plaquette_model}, to arrive at a gauge invariant ansatz whose input is not simply gauge invariant constructs. A schematic of the LGNWF ansatz is shown in Figure~\ref{fig:architecture}.

This approach can be easily generalized to larger lattices. Even after training, the LGNWF can be finetuned on larger lattices requiring only one extra parameter for each new unique distance, $d(x,x')$, introduced by the larger lattice. Furthermore, whilst these will be applied to the spherical representation of $SU(2)$, the LGNWF ansatz can also be applied to any $SU(N)$ LGT.
For more details about these transformations, including various configurations and neural network architectures, and the scaling of this method we refer the reader to Appendix~\ref{sec:model_details} and \ref{app:sec:scaling} respectively.

\subsection{Observables}
To capture the improved performance of the LGNWF ansatz, we define the relative energy change achieved by this ansatz compared to the purely invariant Jastrow ansatz from Eq. \eqref{eq:plaquette_model}, as

\begin{equation}
    \delta E = \frac{E_\text{LGNWF} - E_\text{Jastrow}}{E_\text{Jastrow}}.
    \label{eq:energy_improvement}
\end{equation}

Beyond the energy of the ground state, there remains the question of whether these neural wavefunctions capture other physical properties of the ground state. To answer this question, we consider the scaling of traced plaquettes of size $l \times h$, $\langle W^{l\times h}\rangle$. One expects that in the limit of small $\lambda$ (known often as the \textit{strong} coupling limit) the expectation value of the plaquette should scale with the area of the plaquette, whereas at large $\lambda$ it should scale with its perimeter. The Creutz ratio was introduced in~\cite{creutzRatio} to isolate the amount of area law scaling. The Creutz ratio is defined as

\begin{equation}
    \chi = \frac{1}{N_x}\sum_x-\ln \left(\frac{\langle W^{2\times2}(x)\rangle\langle W^{1\times1}(x)\rangle}{\langle W^{1\times2}(x)\rangle\langle W^{2\times1}(x)\rangle} \right),
\end{equation}
for $N_x$ lattice sites $x$.

\section{Results}

\begin{figure*} 
    \centering
    \includegraphics[width=\linewidth]{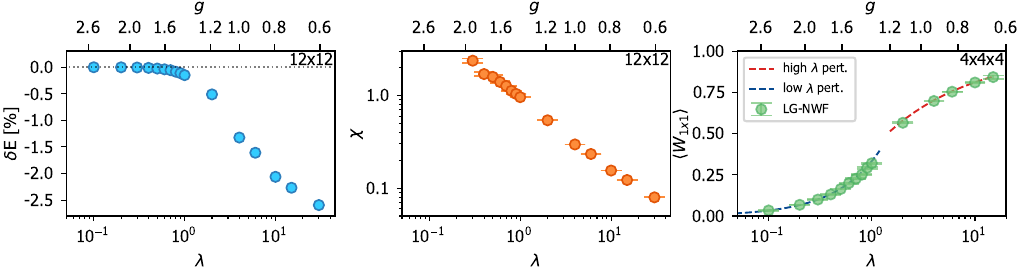}
    \caption{The relative change in energy, defined in Eq.~\eqref{eq:energy_improvement}, achieved using the LGNWF ansatz compared to the two-body Jastrow ansatz (left). Data are shown against $\lambda$ and $g$, where $\lambda = 4/g^4$. The Creutz ratio measured on the ground states found using the LGNWF ansatz (center). The average traced plaquette (divided by 2 for normalization) of the ground states compared to perturbative expansions from~\cite{chin1985exact} (right).
    The Monte Carlo errors for the energy are too small to be distinguished from zero and are thus omitted, whereas for the Creutz ratio and plaquette they are the standard error on the mean across the 144 (64) different lattice sites for the 2- (3-) dimensional lattice.  }
    \label{fig:energy_creutz_wilson}
\end{figure*}

For the following results we performed VMC calculations for a fixed number of iterations, 14,000 (19,000) for the 2- (3-) dimensional lattices. This includes training on smaller lattices and then using these to initialize larger systems. Training was performed using 1,024 samples per iteration, this was increased for the observables reported in this work and will be mentioned alongside the result. For further details see Appendix~\ref{sec:model_details}. 

To establish that the non-local information sharing and increased expressivity facilitated by the LGNWF ansatz improves the variational estimate of the ground state, Figure~\ref{fig:energy_creutz_wilson} (left) shows the energy improvements of the LGNWF ansatz over the Jastrow ansatz for a system of $12\times 12$ links. It is clear that the LGNWF consistently finds a variational state with lower energy and is thus a better representation of the ground state. Furthermore, the improvement in energy increases with increased $\lambda$. Negligible improvement at small $\lambda$ is to be expected; in Ref.~\cite{chin1985exact} the authors show that in the limit of small $\lambda$ the wavefunction should resemble a one-body Jastrow ansatz and therefore the one- and two-body Jastrow ansatz that we are comparing to already contains the optimal solution at small $\lambda$. These data were measured using 65,536 samples and the resulting Monte Carlo (MC) error across this ensemble is too small to be discerned visually, however this error does not account for any uncertainties related to the VMC procedure converging. See Appendix~\ref{sec:model_details} for further discussions on the related uncertainties and the equivalent data for the 3D lattice.

With ground state neural wavefunctions in-hand, what remains is to show that these representations capture more than just a low energy. Figure~\ref{fig:energy_creutz_wilson} (middle) shows the Creutz ratio $\chi$ decreasing monotonically with increasing $\lambda$, consistent with the predictions that the scaling of the plaquette goes from area-dominated to perimeter-dominated. These data were measured using 524,288 samples. Figure~\ref{fig:energy_creutz_wilson} (right) shows the expectation value of the $1\times 1$ traced plaquette averaged over all of the lattice sites for the $4\times 4\times 4$ system using 65,536 samples alongside the perturbative expansions from Ref.~\cite{chin1985exact}. There is clearly very strong agreement between the data and theory in both small and large $\lambda$ regimes (known as the strong and weak coupling expansions respectively). This changeover from the weak coupling to strong coupling behaviors occurs in the range $1 \leq \lambda \leq 2$, in agreement with earlier estimations on similar sized lattices using VMC with a Jastrow ansatz~\cite{chin1985exact}, QMC~\cite{PhysRevD.21.2308, engels1990finite}, and the density of states~\cite{PhysRevLett.109.111601}.

\section{Conclusion and outlook}
In this work we present a flexible yet gauge invariant ansatz for simulating $SU(2)$ lattice gauge theory in the full continuous representation. We employed bespoke gauge equivariant neural networks with a physics-informed gauge invariant final layer to the task of finding the ground state of the $SU(2)$ LGT Hamiltonian. In doing so we highlighted the improvements that can be made with the inclusion of these gauge equivariant networks and showed that measurements taken on these ground states agree well with theoretical predictions. Our combination of various gauge equivariant layers, alongside their inclusion in the framework of VMC, opens up a new avenue to study non-Abelian gauge theories; moreover, these techniques are scalable, applicable to any $SU(N)$ LGT, and facilitate the advancements in condensed matter and quantum information to perforate the study of high-energy physics.

A natural next step is to incorporate dynamical fermionic fields, leveraging recent neural representations of fermionic Hamiltonians~\cite{pescia2024message, smith2024ground, robledo2022fermionic, pfau2024accurate, gu2025solving, chen2025neural, romero2025spectroscopy}. Another is to target low-energy eigenstates~\cite{pfau2024accurate, romero2025spectroscopy} and thermal states~\cite{torlai2018latentspace, nomura2021purifying, nys2024realtime}.
Beyond static properties, our framework is compatible with variational real-time evolution, providing access to non-equilibrium phenomena in gauge theories beyond one spatial dimension. In particular, variational Monte Carlo with expressive neural quantum states has recently emerged as a powerful approach for simulating dynamics in 2D and 3D~\cite{schmitt2020quantum, nys2024ab, van2024many, schmitt2022quantum, gravina2025neural, medvidovic2023variational, sinibaldi2023unbiasing}, including continuous representations~\cite{nys2024ab, medvidovic2023variational}, fermionic degrees of freedom~\cite{nys2024ab}, and thermal state dynamics~\cite{nys2024realtime}. Extending our approach to include these features could shed light on open problems related to such as identifying the mechanisms determining the time scales of the thermalization and the sensitivity of the dynamics to the initial conditions in field theories
~\cite{bauer2023quantum}.

\section{Code and data availability}
The code and data used in this work are openly available at~\cite{spriggs_gitlab, spriggs_zenodo}.

\begin{acknowledgments}
    The authors would like to thank Simon Hands, Patrick Emonts, and Ana Silva for fruitful discussions. 
    The simulations were carried out with a custom-developed code based on \texttt{JAX}~\cite{jax2018github} and \texttt{NetKet}~\cite{vicentini2022netket}. Some of these simulations were performed on the DelftBlue supercomputer~\cite{DHPC2024}.
    This work is part of the project Engineered Topological Quantum Networks (Project No.VI.Veni.212.278) of the research program NWO Talent Programme Veni Science domain 2021 which is financed by the Dutch Research Council (NWO). This publication is also part of the project Optimal Digital-Analog Quantum Circuits with file number NGF.1582.22.026 of the research programme NGF-Quantum Delta NL 2022 which is (partly) financed by the Dutch Research Council (NWO) and the Dutch National Growth Fund initiative Quantum Delta NL. This project was supported by the Kavli Foundation.
\end{acknowledgments}

\newpage
\bibliography{biblio_with_links}

\appendix

\onecolumngrid

\section{Further Hamiltonian details} \label{app:sec:Hamiltonian}
We will now briefly cover some of the details needed for a more intuitive understanding of the Hamiltonian defined in Eq.~\eqref{eq:hamiltonian_abstract_form}, noting in particular the difficulties introduced by gauge transformations. More a broader introduction to LGTs we point the reader to Ref.~\cite{gattringer2009quantum}.

Of the terms in the Hamiltonian, the electric field term will be discussed as a differential operator acting on a polar representation of the links in Appendix~\ref{sec:impl_details}; here we will focus on the plaquette term. Whilst the details of gauge invariance do not depend on the representation, it may help to consider all $SU(2)$ elements as $2\times2$ special unitary matrices for this Section.

\begin{figure}
    \centering
    \includegraphics[width=\textwidth]{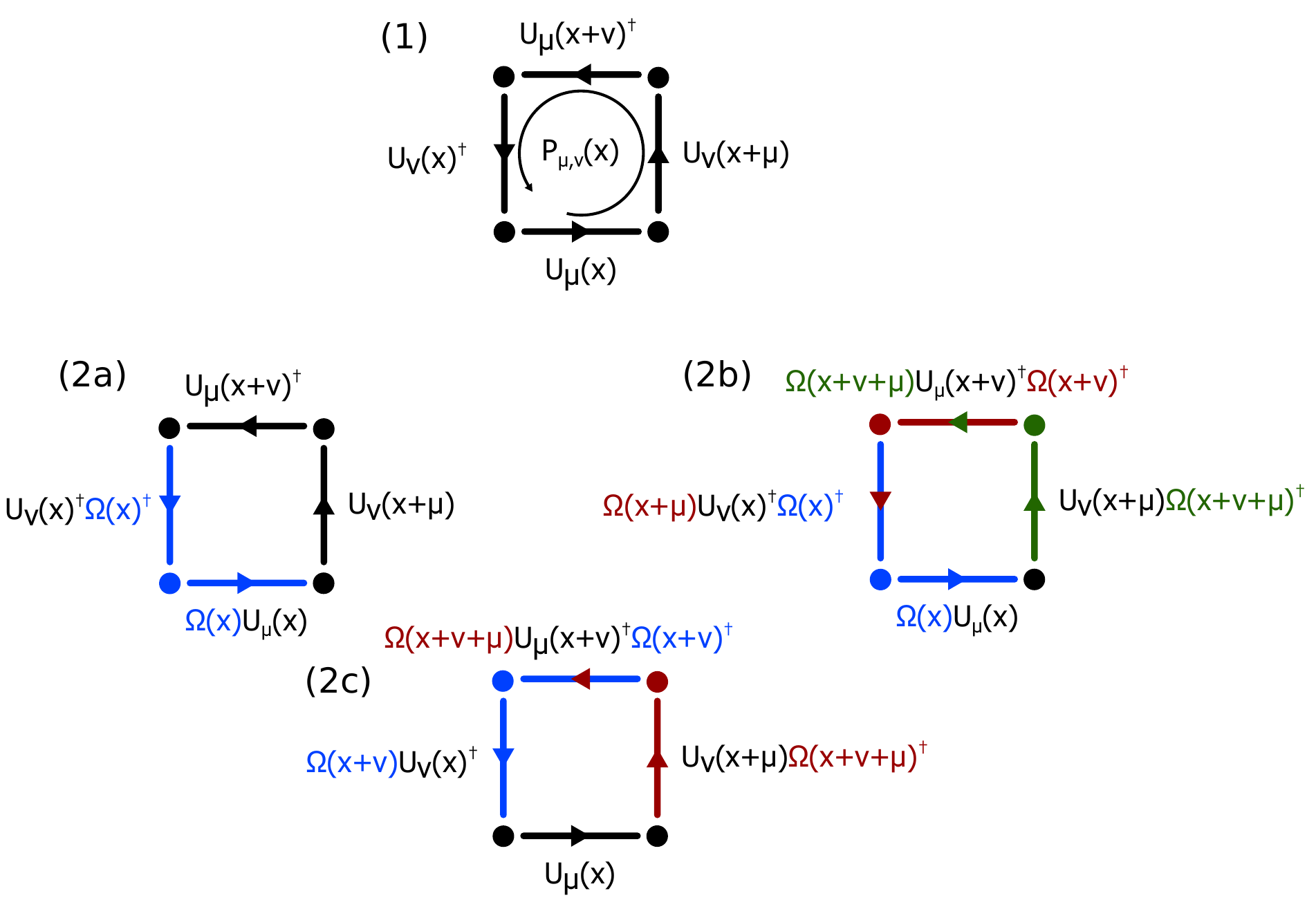}
    \caption{Four \emph{physically-equivalent} configurations that differ by gauge transformation. The unaltered configuration, label (1), also shows the definition of the plaquette: the product of the links forming a closed $1\times1$ loop counter-clockwise, starting at the bottom left. Configurations (2a-c) differ by the application of one or more gauge transformation; the transformation affects all of the links that touch the transformed node and this is represented by the change in color. The transformation introduces local $SU(2)$ elements $\Omega(y)$ on either the left- or right-hand side of the link. Computing the plaquette after the gauge transformation may lead to a different value, but close inspection will show that the trace of the plaquette is unchanged by gauge transformation.}
    \label{fig:app:gauge_T}
\end{figure}

The lattice considered in this work is discrete in space with links connecting nearest neighbor nodes. In a theory without fermions (as we are considering in this work), nothing physical occurs at the nodes, it is the links that contain the physics. An individual link, $U_\mu(x)$, is oriented and points from node $x$ to $x+\mu$; is itself an element of $SU(2)$. To swap the orientation and point from $x+\mu$ to $x$ one can use the Hermitian conjugate of the original link, $U_\mu(x)^\dagger$, this is used for a more compact notation. If any two configurations of links differ only by gauge transformation then they are physically equivalent. In the most general case, a single link changes under gauge transformation following:
\begin{equation}
    U_\mu(x) \rightarrow \Omega(x)U_\mu(x)\Omega(x+\mu)^\dagger.
\end{equation}
Here, $\Omega(x)$ and $\Omega(x+\mu)^\dagger$ are random and, in general, different elements of $SU(2)$. The gauge transformation affects every link that touches a transformed node, but not every node need be transformed. A sketch of three different gauge transformations affecting a configuration is shown in Figure~\ref{fig:app:gauge_T}. As there are infinitely many elements in the continuous gauge group of $SU(2)$, there are infinitely many physically equivalent configurations.

The plaquette is a product of four links forming a closed, counter-clockwise, $1\times1$ loop, beginning at the bottom left node. A plaquette originating at $x$ is defined as
\begin{equation}
    P_{\mu,\nu}(x) = U_\mu(x)U_\nu(x+\mu)U_\mu(x+\nu)^\dagger U_\nu(x)^\dagger.
\end{equation}
Under a general gauge transformation this becomes
\begin{equation}
    \begin{split}
     P_{\mu,\nu}(x) \rightarrow  & \Omega(x)U_\mu(x)\Omega(x+\mu)^\dagger\\
     &\Omega(x+\mu)U_\nu(x+\mu)\Omega(x+\mu+\nu)^\dagger\\
     &\Omega(x+\mu+\nu)U_\mu(x+\nu)^\dagger\Omega(x+\nu)^\dagger\\
     &\Omega(x+\nu)U_\nu(x)^\dagger\Omega(x)^\dagger\\
     &= \Omega(x)P_{\mu,\nu}(x)\Omega(x)^\dagger.
    \end{split}
\end{equation}
The plaquette is, in general, not gauge invariant. On the contrary, given the cyclic nature of the trace it is clear to see that the \emph{trace} of the plaquette is invariant under gauge transformation. This is why the plaquette term in the Hamiltonian is unchanged by gauge transformations.

In all instances throughout this work we will be using a rescaled version of the Hamiltonian with the factor $\frac{g^2}{a_s}$ divided out. In the figures of the main text this prefactor always cancels and thus does not have an affect.

\section{Practical implementation}\label{sec:impl_details}
In this Section we detail the VMC implementation. We represent the link $SU(2)$ matrices with 4 real numbers $[a^0, a^1, a^2, a^3] \in \mathbb{R}^4$
\begin{align}\label{eq:represenation_R4}
    &U = a^0 I_{2 \times 2} + i(a^1 \sigma^{x} + a^2 \sigma^{y} + a^3 \sigma^{z}), \\
    &\sum_k (a^k)^2 = 1, 
\end{align}
where $\sigma^{x}, \sigma^{y}, \sigma^{z}$ are the $2 \times 2$ Pauli matrices
\begin{align}
\sigma^{x}=\begin{pmatrix}0&1\\1&0\end{pmatrix}, \\
\sigma^{y}=\begin{pmatrix}0&-i\\i&0\end{pmatrix}, \\
\sigma^{z}=\begin{pmatrix}1&0\\0&-1\end{pmatrix}.
\end{align}
This representation allows for faster matrix multiplications and traces. Hilbert space samples are therefore represented by $N_e \times 4$ real values, with $N_e$ the number of edges.

For Markov chain Monte Carlo (MCMC) sampling, we define the following random-walk update with scale $\epsilon$ in the Lie algebra
\begin{align}
u &= \mathcal{N}(0, I_3), \\
\omega &= \epsilon u, \\
V &= \exp\left(-i\sum_{a=x,y,z}\sigma^a\frac{\omega^a}{2} \right),
\end{align}
where we defined the exponential map from the Lie algebra $\omega^a$ to the Lie group elements $V$. Notice that for $\epsilon \to 0$ we get $V \to I_{2\times 2}$. We then generate updates of the link matrices via
\begin{align}
    U \to V U.
\end{align}

We vary the scale $\epsilon$ in the sampler dynamically to achieve a targeted acceptance rate of approximately $A_{\textrm{tgt}} = 57\%$. Therefore, we use the update rule 
\begin{align}
    \epsilon \leftarrow \epsilon \frac{\max(A, 0.05)}{A_{\textrm{tgt}}},
\end{align}
based on the current acceptance $A$~\cite{schatzle2023deepqmc}.

Our Hamiltonian contains the Laplace-Beltrami operator on $S^3$. In spherical coordinates this reads~\cite{chin1985exact}
\begin{align}\label{eq:laplace_beltrami}
    \nabla^2_{S^3} &= \frac{1}{\sin^2(\frac{\rho}{2})}\left(\frac{\partial}{\partial \rho}\left(4\sin^2\left(\frac{\rho}{2}\right)\frac{\partial}{\partial \rho}\right) + \frac{1}{\sin\theta}\frac{\partial}{\partial \theta}\left(\sin\theta \frac{\partial}{\partial\theta}\right) + \frac{1}{\sin^2\theta}\frac{\partial^2}{\partial \phi^2} \right),
\end{align}
which we implement as follows. Since we parametrize $\log \Psi$ and apply the above operator to each individual link $e$, we have, for the local operator, terms
\begin{align}
    \sum_e \frac{\nabla^2_{S^3, e} \Psi(\vec{U})}{\Psi(\vec{U})} &= \sum_e  \left[\nabla^2_{S^3, e} \log \Psi(\vec{U}) + (\nabla_{S^3, e} \log \Psi(\vec{U}))^2\right],
\end{align}
where
\begin{align}
    \abs{\nabla_{S^3} \log \Psi}^2 &= 4 \left[ \left(\frac{\partial f}{\partial \rho}\right)^2 +  \left(\frac{1}{2\sin \frac{\rho}{2}}\frac{\partial f}{\partial \theta}\right)^2 + \left(\frac{1}{2\sin \frac{\rho}{2} \sin \theta}\frac{\partial f}{\partial \phi}\right)^2  \right] .
\end{align}
In the practical implementation, we use automatic differentiation to compute the Laplacian, where we differentiate with respect to the spherical coordinates through both the wave-function representation and the spherical representation $U(\rho, \theta, \phi)$ in Eq.~\eqref{eq:link_ito_polars}.

We can then perform VMC in the standard way~\cite{vicentini2022netket} using samples $\vec{U}$ containing $SU(2)$ matrices on all the edges of the lattice. We use stochastic reconfiguration~\cite{sorella1998green} with a diagonal regularization for stability. The diagonal shift used was $10^{-4}$ and the learning rate scheduler followed the convention of~\cite{smith2018superconvergencefasttrainingneural}, with a learning rate warm-up followed by a cosine decay. 1,024 samples were used per iteration for both 2D and 3D.

\section{Conventions leading to the Laplace-Beltrami operator} \label{app:sec:Lapl_belt}
In Eq.~\eqref{eq:represenation_R4}, we represent $SU(2)$ matrices as an $S^3$ sphere in $\mathbb{R}^4$. Matching with  Eq.~\eqref{eq:link_ito_polars}, we have the hyperspherical representation
\begin{align}
    a^0 &= \cos \frac{\rho}{2}, \\
    a^k &= -\sin \frac{\rho}{2} n^k, \\
\end{align}
where
\begin{align}
    n^k=[\sin\theta\cos\phi,\;\sin\theta\sin\phi,\;\cos\theta].
\end{align}
The corresponding differentials read
\begin{align}
    \dd a^0 &= -\tfrac12\,\sin(\tfrac{\rho}{2})\,\dd\rho, \\
    \dd a^k &= - \tfrac12\,\cos(\tfrac{\rho}{2})\,n^k\,\dd\rho - \sin(\tfrac{\rho}{2})\,\dd n^k,
\end{align}
where each $\dd n^k$ is given by
\begin{align}
    \dd n^1 &= \cos\theta\cos\phi\,\dd\theta - \sin\theta\sin\phi\,\dd\phi, \\
    \dd n^2 &= \cos\theta\sin\phi\,\dd\theta + \sin\theta\cos\phi\,\dd\phi, \\
    \dd n^3 &= -\sin\theta\,\dd\theta.
\end{align}

The metric on $\mathbb{R}^4$ yields the unit-radius $S^3$ metric
\begin{align}
    \dd s^2 &= \sum_{i=0}^4 \dd (a^i)^2,\\
    &= \frac{1}{4}  \dd \rho^2 + \sin^2 (\frac{\rho}{2}) \sum_k \dd (n^k)^2, \\
    &= \frac{1}{4}  \dd \rho^2 + \sin^2 (\frac{\rho}{2}) \left( \dd \theta^2 + \sin \theta \dd \phi^2 \right), \\
    &= \frac{1}{4} \left[\dd \rho^2 + 4 \sin^2 (\frac{\rho}{2}) \left( \dd \theta^2 + \sin^2 \theta \dd \phi^2 \right) \right],
\end{align}
where we have exploited the relations
\begin{align}
\sum_k (n^k)^2 = 1, \\
\sum_k n^k \dd n^k = 0.
\end{align}
Hence, for the metric $g$, we obtain
\begin{align}
g^{ii} &=  \left[ \frac{1}{4}, \,  \sin^2(\frac{\rho}{2}), \, \sin^2(\frac{\rho}{2})\sin^2\theta  \right], \\
\sqrt{\det g} &= \frac{1}{2} \sin^2 (\frac{\rho}{2}) \sin \theta,\\
g_{ij} &= (g^{-1})^{ij}.
\end{align}
These can be inserted into the general definition of the Laplace Beltrami operator
\begin{align}
\nabla_{S^3}^2 &= \frac{1}{\sqrt{\det g}} \partial^i \left(\sqrt{\det g} g_{ij} \partial^j \right), \\
&= \frac{1}{\sqrt{\det g}} \partial^i \left(\sqrt{\det g} (g^{ii})^{-1} \partial^i \right), \\
&=  \frac{4}{\sin^2 \frac{\rho}{2}} \frac{\partial }{\partial \rho}\left(\sin^2 \frac{\rho}{2} \frac{\partial }{\partial \rho}\right) +
\frac{1}{\sin^2 \frac{\rho}{2}} \frac{1}{\sin\theta} \frac{\partial}{\partial \theta} \left( \sin \theta \frac{\partial}{\partial \theta} \right)  
+ \frac{1}{\sin^2 (\frac{\rho}{2}) \sin^2\theta} \frac{\partial^2}{\partial \phi^2}, \\
&= \frac{1}{\sin^2 \frac{\rho}{2}} \left[\frac{\partial }{\partial \rho}\left(4\sin^2 \frac{\rho}{2} \frac{\partial }{\partial \rho}\right) +
 \frac{1}{\sin\theta} \frac{\partial}{\partial \theta} \left( \sin \theta \frac{\partial}{\partial \theta} \right)  
+ \frac{1}{\sin^2\theta} \frac{\partial^2}{\partial \phi^2} \right], \\
&= 4 \left(
\frac{\partial^2}{\partial \rho^2} + \cot (\frac{\rho}{2}) \frac{\partial}{\partial \rho} + \frac{1}{4\sin^2 (\frac{\rho}{2})} \left[
\frac{\partial^2}{\partial \theta^2} + \cot \theta \frac{\partial}{\partial \theta} +  \frac{1}{\sin^2 \theta}\frac{\partial^2}{\partial \phi^2}
\right]
\right),
\end{align}
which can be computed directly from the diagonal of the Hessian and the gradient.
Hence, we obtain the Laplace Beltrami operator in Eq.~\eqref{eq:EE}.
From a similar treatment, we can also obtain the square of the gradient
\begin{align}
    \abs{\nabla f}^2 &= g_{ij} \partial^i f \partial^j f, \\
    &= (g_{ii})^{-1} (\partial^i f)^2, \\
    &= 4 \left(\frac{\partial f}{\partial \rho}\right)^2 + \frac{1}{\sin^2 \frac{\rho}{2}} \left(\frac{\partial f}{\partial \theta}\right)^2 + +\frac{1}{\sin^2 \frac{\rho}{2} \sin^2 \theta}\left(\frac{\partial f}{\partial \phi}\right)^2, \\
    &= 4 \left[ \left(\frac{\partial f}{\partial \rho}\right)^2 +  \left(\frac{1}{2\sin \frac{\rho}{2}}\frac{\partial f}{\partial \theta}\right)^2 + \left(\frac{1}{2\sin \frac{\rho}{2} \sin \theta}\frac{\partial f}{\partial \phi}\right)^2  \right].
\end{align}

These expressions feed directly into the local kinetic energy 
\begin{align}
    \frac{\nabla^2 \Psi}{\Psi} &= \Delta\log\Psi + |\nabla\log\Psi|^2,
\end{align}
that are used in variational Monte Carlo.

\section{Variational model}\label{sec:model_details}

\subsection{Equivariance}
The lattice gauge neural wavefunction shown in this work is constructed from the following building blocks: two kinds of gauge equivariant transformations (the plaquette equivariant layer of Eqs.~\eqref{eq:link_equivariant_equations-1}-\eqref{eq:link_equivariant_equations-2} and the node equivariant layer from Eqs.~\eqref{eq:node_equivariant_equations-1}-\eqref{eq:node_equivariant_equations-2}) followed by the Jastrow plaquette model from Eq.~\eqref{eq:plaquette_model}. We will first establish the equivariance of the two layers, and then elucidate the final choice of the ansatz.

We will begin with the discussion the equivariance of the plaquette equivariant transformation in Eqs.~\eqref{eq:link_equivariant_equations-1}-\eqref{eq:link_equivariant_equations-2}. To start, we re-establish that the link transforms under a change of gauge through
\begin{equation}\label{eq:appendix:gauge_on_U}
    \mathcal{T}_\Omega U_{\mu}(x) = \Omega(x) U_\mu(x) \Omega(x+\mu)^\dagger.
\end{equation}
Now that this has been stated, we will show the result of applying the plaquette transformation layer on the gauge-transformed link. To begin with, we show that the plaquette $P_{\mu, \nu}(x)$ transforms under $\mathcal{T}_\Omega$ as
\begin{align}\label{eq:appendix:gauge_on_p}
    \mathcal{T}_\Omega P_{\mu, \nu}(x) &= \Omega(x) P_{\mu, \nu}(x) \Omega(x)^\dagger, \\
    &= (\Omega(x) V) e^{i\lambda} ( \Omega(x)V)^\dagger,
\end{align}
where $\lambda = [\lambda_1, \lambda_2]$ is a vector of the sorted eigenvalues of $P$ and $V$ are the corresponding eigenvectors. (Note that for $SU(N)$ theories for $N>2$ one has to do more work to impose a canonical ordering of eigenvalues, see~\cite{boyda2021sampling} for applications to $SU(3)$.) We can also note that the updated plaquette, $P_{\mu, \nu}'(x) = Ve^{i\lambda'} V^\dagger$, transforms the same way:
\begin{align}
    \mathcal{T}_\Omega P_{\mu, \nu}'(x) &= (\Omega(x) V) e^{i\lambda'} ( \Omega(x)V)^\dagger, \\
    &= \Omega(x) P_{\mu, \nu}'(x) \Omega(x)^\dagger,
\label{eq:appendix:gauge_on_p'}
\end{align}
if and only if the neural network that computes $\lambda'$ is invariant under gauge transformation: trivially satisfied by acting on gauge invariant information like the traced plaquettes.
Combining Eq.~\eqref{eq:link_equivariant_equations-2} with Eqs.~\eqref{eq:appendix:gauge_on_U}-\eqref{eq:appendix:gauge_on_p'} to compute the action of the plaquette transformation layer on gauge-transformed links we get
\begin{align}
    [ \mathcal{T}_\Omega U_\mu(x)]' = & \left[ \Omega(x) P_{\mu, \nu}'(x) \Omega(x)^\dagger \right] \left[ \Omega(x) P_{\mu, \nu}(x) \Omega(x)^\dagger \right]^\dagger \left[ \Omega(x) U_\mu(x) \Omega(x+\mu)^\dagger \right], \\
    = &\; \Omega(x) \left[P_{\mu, \nu}'(x) P_{\mu, \nu}(x)^\dagger U_\mu(x)\right] \Omega(x+\mu)^\dagger, \\
    = &\; \Omega(x) U_\mu'(x) \Omega(x+\mu)^\dagger, \\
    = &\; \mathcal{T}_\Omega U_\mu'(x).\\
\end{align}
Thus demonstrating that the plaquette equivariant layer is equivariant with respect to a gauge transformation.

Likewise, for the node equivariant layer defined in Eqs.~\eqref{eq:node_equivariant_equations-1}-\eqref{eq:node_equivariant_equations-2} we can show equivariance by following the work of~\cite{favoni2025symmetrypreservingneuralnetworkslattice}. We first notice that each element in $C_{\mu, \nu}^i(x)$ transforms as
\begin{align}
    \mathcal{T}_\Omega C_{\mu, \nu}^i(x) = \Omega(x)C_{\mu, \nu}^i(x)\Omega(x)^\dagger.
\end{align}
This means that $\epsilon_\mu(x)$ transforms as
\begin{align}
    \mathcal{T}_\Omega \epsilon_\mu(x) &= e^{i\sum_{i} \gamma_{\mu,i} \Omega(x)[C_{\mu, \nu}^i(x)]_{aH} \Omega(x)^\dagger}, \\
    &=\sum_{n=0}^{\infty}\frac{1}{n!}\left( i\sum_{i} \gamma_{\mu,i} \Omega(x)[C_{\mu, \nu}^i(x)]_{aH} \Omega(x)^\dagger\right)^n,\\    &=\Omega(x)\sum_{n=0}^{\infty}\frac{1}{n!}\left( i\sum_{i} \gamma_{\mu,i} [C_{\mu, \nu}^i(x)]_{aH} \right)^n\Omega(x)^\dagger,\\
    &=\Omega(x) e^{i\sum_{i} \gamma_{\mu,i} [C_{\mu, \nu}^i(x)]_{aH} }\Omega(x)^\dagger, \\
    &= \Omega(x)\epsilon_\mu(x)\Omega(x)^\dagger.
\end{align}
Together with Eq.~\eqref{eq:node_equivariant_equations-2}, we then get
\begin{align}
     \left[\mathcal{T}_\Omega U_\mu(x)\right]'&= \left[\Omega(x) \epsilon_\mu(x) \Omega(x)^\dagger\right] \left[\Omega(x) U_\mu(x) \Omega(x+\mu)\right], \\
    &= \Omega(x) \left[\epsilon_\mu(x)  U_\mu(x) \right]\Omega(x+\mu), \\
    &= \Omega(x) U'_\mu(x)\Omega(x+\mu), \\
    &= \mathcal{T}_\Omega U'_\mu(x),
\end{align}
demonstrating the equivariance.

\subsection{Model details}
For simplicity of the explanation, throughout this work we have been discussing the update layers as if they only act on a single link; in practice we update many links in a single layer and often do so in parallel. All links can be acted on in parallel during the node equivariant layer, and all of the required plaquettes can be computed only once per layer. For the plaquette equivariant layer, however, we chose to update all of the links oriented in the same dimension at the same time, but only on dimension per plane. The plaquettes are recomputed for each plane due to the updated links from the previous plane. This is sufficient to ensure that all of the plaquettes after the equivariant layer have been updated using the equivariant transform without having to update every link. More complex patterns of updates can be performed, like in~\cite{boyda2021sampling}, however these are constructed because the authors require the equivariant layer to also be invertible - not a constraint that is relevant here.

\begin{figure}
    \centering
    \includegraphics[width=0.5\linewidth]{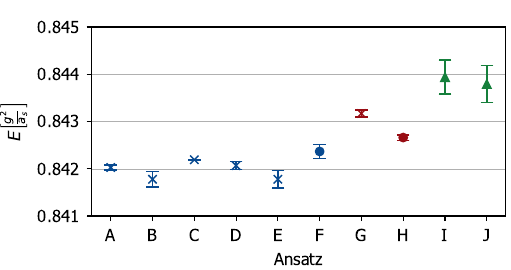}
    \caption{Variational ground state energy for a range of ansatze outlined in Table~\ref{tab:app:different_ansatze}. Errors are the Monte Carlo error of the energy over the final set of samples.}
    \label{fig:app:different_anstaze}
\end{figure}

There still remains some freedom when combining the equivariant layers, as well as the configuration of the layers themselves. The node equivariant layer can use any closed loops as the elements $C_{\mu, \nu}^i(x)$, however, for computational efficiency we chose the smallest closed loops: plaquettes. As for the plaquette equivariant layer, the eigenvalue of the plaquette $\lambda$ can be updated in any gauge invariant way: in this work we limited this to a neural network acting taking as input all of the traced plaquettes on the lattice in the same plane as the link being updated. We tested two architectures: a convolutional neural network (CNN) and vision transformer (ViT) with various hyperparameters. Beyond this, one can choose to use multiple repeated blocks of node and plaquette equivariant layers, only plaquette equivariant layers, or no equivariance at all. Finally, one can also replace the final Jastrow model with a fully connected network (FCN) acting on the traced plaquettes. A comparison between the ground state energy of a selection of representative ansatze on a $4\times 4$ lattice are shown in Figure~\ref{fig:app:different_anstaze} and detailed in Table~\ref{tab:app:different_ansatze}. As we find the ground state to the rescaled version of the Hamiltonian in Eq.~\eqref{eq:hamiltonian_abstract_form} the energy is reported in units of $\frac{g^2}{a_s}$. These variational states were trained using $1,024$ samples per iteration for $4,000$ iterations. From these data, one notable pattern arises: from the separation between the blue, red, and green data (comprised of plaquette and node equivariant layers, only plaquette equivariant layers, and no equivariance, respectively) it would seem that the choice of equivariant layer(s) matters more than the specifics of the layer implementations themselves. Particularly for the ansatze comprised of both types of equivariant layer, the choice of neural network within the plaquette equivariant layer does not offer much of a performance increase. It should be noted, however, that the errors shown are just the statistical errors arising from Monte Carlo sampling of the energy during the final iteration of VMC; these errors do not capture any variation between repeated runs of VMC nor the fluctuation over training. This is significant as particularly ansatze containing the FCN were less stable than those with the Jastrow final layer; this is possibly due to initialization. The one-body term of the Jastrow ansatz was initialized with the optimized values from~\cite{chin1985exact}, and the two-body term was initialized to be small; the FCN was initialized using the \texttt{lecun\_normal} initializer from \texttt{JAX} which was itself the most stable of commonly used strategies.

\begin{table}
    \centering
\begin{tabular}{|c|c|c|c|c|c|c|}
\hline
\textbf{Figure~\ref{fig:app:different_anstaze} label} & \multicolumn{3}{c|}{\textbf{Plaquette layer}} & \textbf{Node layer}& \textbf{Num. equivariant blocks} & \textbf{Final layer} \\
\cline{2-4}
                       & \textbf{Architecture} & \textbf{Num. layers} & \textbf{Other} & & & \\
\hline
A                 & CNN             & 1             & K=3  & Yes & 1 & Jastrow \\
B                 & CNN             & 1             & K=1  & Yes & 1 & Jastrow \\
C                 & CNN             & 2             & K=3  & Yes & 1 & Jastrow \\
D                 & CNN             & 1             & K=3  & Yes & 2 & Jastrow \\
E                 & CNN             & 1             & K=3  & Yes & 1 & FCN \\
F                 & ViT             & 1             & NH=2, D=4 & Yes & 1 & Jastrow \\
G                 & CNN             & 1             & K=3  & No & 1 & Jastrow \\
H                 & ViT             & 1             & NH=2, D=4 & No & 1 & Jastrow \\
I                 & None             & N/A             & N/A  & No & 0 & Jastrow \\
J                 & None             & N/A             & N/A  & No & 0 & FCN \\
\hline
    \end{tabular}
    \caption{The selection of different ansatz configurations shown in Figure~\ref{fig:app:different_anstaze}. The top row was used throughout this work as the LGNWF. Shorthands used: CNN = convolutional neural network, ViT = vision transformer, K = kernel size for the CNN, NH = number of heads in the ViT, and D = embedding dimension of the ViT.}
    \label{tab:app:different_ansatze}
\end{table}

\subsection{Initialization procedure}\label{app:sec:init}
\begin{figure}
    \centering
    \includegraphics[width=0.5\linewidth]{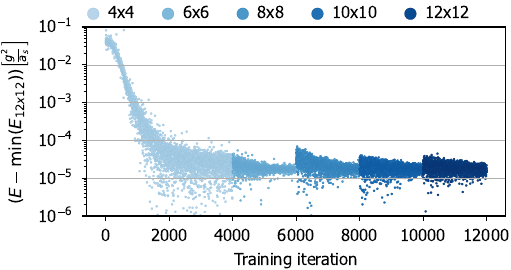}
    \caption{The energy of each sample in the Markov chains during VMC training with relative to the lowest energy found on the $12\times 12$ lattice. The ansatz used for this training was the LGNWF used throughout the main text with $\lambda=0.1$.}
    \label{fig:app:training}
\end{figure}

Given that the computational cost of VMC increases with system size, we chose to train the network on progressively larger lattices. For the 2D case, the network was first trained from scratch on a $4\times4$ lattice, and then these parameters were carried over and further trained on $6\times 6$, $8\times 8$, $10\times 10$, and then finally $12\times 12$ lattices. For the 3D case, we trained on $3\times 3\times 3$ and then moved to $4\times 4\times 4$. Considering the ansatz shown in the main text, A from Table~\ref{tab:app:different_ansatze}, the only parameters that need to be added to work on a larger system is those in the two-body term of the Jastrow layer, $\beta_{ d(x, x')}^{\mu, \nu, \mu', \nu'}$ from Eq.~\eqref{eq:plaquette_model}, for the new distances $d(x, x')$: these were simply initialized to be small. Importantly, the samples from smaller systems cannot be carried over to the larger systems so the Markov chains must re-thermalize. An example of the this training procedure, for $\lambda=0.1$ in 2D, is shown in Figure~\ref{fig:app:training}. From this Figure one can see that the effects of increasing the system size are mild, albeit after some thermalization period.

Figure~\ref{fig:app:training} shows a 12,000 iteration training procedure (with 1,024 samples per iteration) minimizing the energy of a variational state under the Hamiltonian defined in Eq.~\ref{eq:hamiltonian_abstract_form} with the Laplace-Beltrami operator defined in Appendix~\ref{sec:impl_details}. However, for the actual VMC runs that led to the main results of this work the variational states were first trained to minimize the energy of a Hamiltonian with a slightly different Laplace-Beltrami implementation. For the 2D (3D) simulations we trained under the alternative Hamiltonian for 12,000 (17,000) iterations and then both were fine-tuned for a further 2,000 iterations on the version outlined in Appendix~\ref{sec:impl_details}. We compared the final energy of Figure~\ref{fig:app:training} with the version trained under the alternative Hamiltonian and then fine-tuned and they agree within errors.

\begin{figure}
    \centering
    \includegraphics[width=0.5\linewidth]{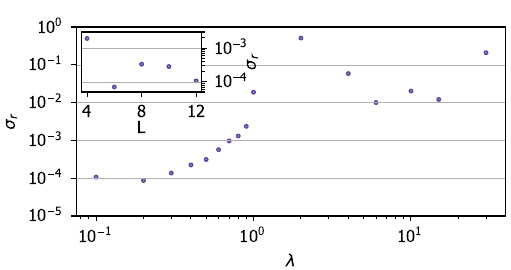}
    \caption{The rescaled variance of the energy, as in Eq.~\eqref{eq:app:rescaled_variance}, averaged over the final 20 iterations of VMC. The main panel shows results for $12\times 12$ lattices at various values of $\lambda$; the inset shows the rescaled variance but against system size ($L \times L$) with fixed $\lambda=0.1$. }
    \label{fig:app:rescaled_variance}
\end{figure}

During the VMC procedure we also compute the energy variance $\sigma$, defined as
\begin{equation}
    \sigma = \frac{\langle \Psi|H^2|\Psi\rangle}{\langle \Psi|\Psi\rangle} - \left(\frac{\langle \Psi|H|\Psi\rangle}{\langle \Psi|\Psi\rangle} \right)^2.
\end{equation}
The variance thus measures how far the variational state $\Psi$ is from being an eigenstate of the Hamiltonian. If one assumes that the nearest eigenstate is the ground state then this can be seen as a measure of the quality of the variational estimate of the ground state. In Ref.~\cite{wu2024variational} the authors suggest rescaling the variance to provide a Hamiltonian-agnostic measure of variational complexity by removing scaling effects of energy and system size. Whilst we do not present the exact ``V-score'' as prescribed in the aforementioned work (due to the difficulty in accessing the ``zero point energy''), we report the rescaled variance, $\sigma_r$ as
\begin{equation}
    \sigma_r = \frac{\sigma N_e}{E^2},
    \label{eq:app:rescaled_variance}
\end{equation}
for $N_e$ edges and energy $E$. The rescaled variance is shown against $\lambda$ and system size for an $L\times L$ system in Figure~\ref{fig:app:rescaled_variance}. As the rescaling allows for comparisons between variational states at different $\lambda$, we can see that finding the ground state becomes increasingly difficult with increasing $\lambda$ until roughly saturating for $\lambda < 1$.

\subsection{3-dimensional energy improvements}
\begin{figure}
    \centering
    \includegraphics[width=0.5\linewidth]{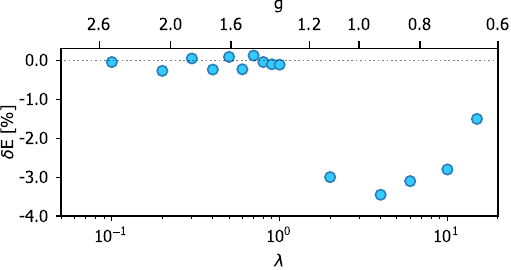}
    \caption{Energy improvements of the LGNWF ansatz as defined in Eq.~\eqref{eq:energy_improvement} for the $4\times4\times4$ lattice. The MC errors over the ensemble of samples would be too small to be clear and are thus omitted.}
    \label{app:fig:3D_energy}
\end{figure}

In Figure~\ref{fig:energy_creutz_wilson} we showed the energy improvement achieved by the LGNWF ansatz in 2D. Figure~\ref{app:fig:3D_energy} shows the equivalent data for the $4\times4\times4$ lattice using 65,536 samples. Clearly, the improvement is less broadly clear; specifically at intermediate $\lambda$ the data fluctuate around the dashed line corresponding to no improvement. At large $\lambda$, however, it is clear that the large improvement that the LGNWF offers is present in 3D, too. We attribute these fluctuations to the neural network having possibly not converged when the measurements were performed. We continue this discussion in Appendix~\ref{app:sec:opt_details} after discussion the optimization details more generally. 

\subsection{Optimization details}\label{app:sec:opt_details}

From Figure~\ref{fig:app:training} one can see that the training occurs for a fixed number of iterations. This number was decided after observing the variance during many VMC trainings over various $\lambda$s and system sizes, but is not formally from any variance- or energy-based stopping criterion. Several alternatives exist: stopping when the energy (or variance) stops changing over a certain number of iterations, stopping when a certain variance threshold has been met, or slowly decreasing the learning rate to 0. Due to the differences in the rescaled variances across the range of $\lambda$ shown in Figure~\ref{fig:app:rescaled_variance} and the computational costs associated with further training with a decaying learning rate, we opted simply for a fixed number of iterations. This is not without complications, however. Particularly in the small $\lambda$ range for the $4\times4\times4$ simulations, the energy estimation for the LGNWF ansatz and the Jastrow ansatz are still fluctuating when training is stopped. During the VMC process only 1024 samples are used to measure the energy, and then 65,536 samples are used to measure the final energies used in Figures~\ref{fig:energy_creutz_wilson} (left) and \ref{app:fig:3D_energy}. Therefore, the fluctuations arising from which point one decides to stop training (especially given the larger MC error associated with the estimate of the energy from only 1,024 samples) outweigh the MC error when the final energy is measured with a factor of 64 times more samples. Estimating the error arising from the stopping point would require measuring the energy on several ground states using the increased number of samples and thus would add significant computational cost. This early stopping is likely the reason for the fluctuations at small $\lambda$ in Figure~\ref{app:fig:3D_energy}.

\section{Scaling of the method} \label{app:sec:scaling}

The method covered in this work could be extended to explore time evolution and non-equilibrium phenomena of LGTs. There are, however, more immediate ways to scale this method: studying larger systems and other $SU(N)$ gauge groups. We will now discuss these briefly. 

The most computationally expensive simulations performed in this work were those of the $12\times12$ and $4\times4\times4$ lattices. The VMC training was performed on a single NVIDIA Tesla A100 80GB GPU and then one NVIDIA Hopper H100 94GB GPU was used for measuring observables. In both the two- and three-dimensional cases, however, we were able to process the samples in parallel (with a batch size of 64) to reduce runtime at the expense of increased memory consumption. This parallelization could be reduced to account for working on larger lattices with a limited amount of GPU memory. Furthermore, these methods can be performed on multi-GPU setups for even larger memory allowances~\cite{vicentini2022netket}. Both of these techniques (decreasing batch size and using multiple GPUs) would allow for the simulation of much larger lattices than shown in this work. 

The most memory-intensive computations are the Laplace-Beltrami operator applied to our network (currently about 85\% of the runtime). These also require small batch sizes, and restrict the parallelism.
Using recent developments on forward-Laplacians~\cite{li2024computational, gao2023folx} can further help reduce the memory and runtime requirements, and increase the parallelism of our approach. Another approach is to combine these improvements with locality-enforcing and sparsity neural architectures, as explored in Refs.~\cite{li2024computational, scherbela2025accurate} in the context of accurate ab-initio neural-network solutions to large-scale electronic structure problems in continuous space.

The second bottleneck at the system sizes studied in this work is efficient decorrelated sampling (about 15\% of the runtime). To address this challenge for larger systems one can introduce methods such as hybrid Monte Carlo~\cite{medvidovic2023variational} and Metropolis Adapted Langevin samplers~\cite{nys2024ab}, that decrease the correlation time compared to the random-walk Monte Carlo used in this work. 

The final bottleneck for larger systems is solving the stochastic reconfiguration linear system~\cite{sorella1998green}, scaling cubic in the number of parameters. Here, recent approaches based on the Woodbury identity can bring this scaling down to linear in the parameter number~\cite{chen2024empowering}, and cubic in the number of samples (which is kept constant across system sizes in general).

The scaling of the ansatz itself for larger systems is rather inexpensive. As mentioned in Appendix~\ref{app:sec:init}, when the specific form of the LGNWF utilizes the CNN in the plaquette equivariant layer then the only new parameters that need to be added to scale to a larger system are those needed to populate $\beta_{ d(x, x')}^{\mu, \nu, \mu', \nu'}$ from Eq.~\eqref{eq:plaquette_model} for the new distances $d(x, x')$. This amounts to only one extra parameter per new distance on the larger lattice. The CNN contains the same number of parameters for any system size. If a ViT (or any other non-local) architecture is used then the scaling of the ansatz will have to account for this. There are, however, efficient implementations of the ViT with improved scaling over conventional versions~\cite{Rende_2025}. 
However, for a system with periodic boundary conditions and with interactions that become negligible beyond a certain distance (\textit{i.e.} the case where the wavefunction is largely determined by local features), then the exact same model can be applied to increasing system sizes. This means that we can tackle larger system sizes without adding additional parameters and with a computational cost scaling linearly in system size.

As for the extension to other gauge groups, from a computational perspective, this is also feasible. For $SU(2)$, one needs three real numbers to uniquely define the group element; in, for example, $SU(3)$, this only rises to eight (and $N^2-1$ in general). We opted for four numbers per link for $SU(2)$ as that parameterization led to faster matrix multiplications. Regardless, the memory usage scales polynomially in $N$, meaning that scaling to higher $N$ theories is within reach. It is also worth noting that much of the attention of QMC calculations is placed on $SU(2)$ and $SU(3)$, rather than significantly larger $N$.

\end{document}